\documentclass{iopart}
\usepackage{graphicx}
\usepackage{amssymb}
\usepackage[normalem]{ulem}
\usepackage{color}

\begin{document}

\title{
Directed Motion of Liquid Crystal Skyrmions With Oscillating Fields 
} 
\author{
A. Duzgun, C. Nisoli, C. J. O. Reichhardt, and  C. Reichhardt 
} 
\address{
Theoretical Division and Center for Nonlinear Studies,
Los Alamos National Laboratory, Los Alamos, New Mexico 87545, USA\\ 
}
\ead{cjrx@lanl.gov}

\begin{abstract}
Using continuum simulations, we show that under
a sinusoidal
electric field,
liquid crystal skyrmions undergo periodic shape oscillations
which produce
controlled directed motion.
The speed of the skyrmion is non-monotonic in the frequency of the applied field,
and exhibits
multiple reversals of the motion
as a function of changing frequency.
We map out the
dynamical regime diagram
of the forward and reverse motion for
two superimposed ac driving frequencies, and show that 
the reversals and directed motion can occur even when only
a single ac driving frequency is present.
Using pulsed ac driving,
we
demonstrate
that the  motion arises due to
an asymmetry in the relaxation times of the skyrmion shape. We 
discuss the connection between our results and ratchet effects observed
in systems without asymmetric substrates.
\end{abstract}

\maketitle

\vskip2pc

\section{Introduction}
Skyrmions are particle like textures that arise in chiral  magnets
\cite{Rossler06,Muhlbauer09,Yu10,Nagaosa13} 
and chiral liquid crystal systems
\cite{Fukuda11,Ackerman14,Leonov14,Nych17,Ackerman17,Duzgun18,Sohn19a}.
When either confinement or strong electric fields
are present,
the liquid crystal (LC) system can
form a uniform nematic, a  cholesteric stripe, or a meron lattice. 
Recently, experiments on LC skyrmions or baby skyrmions
revealed directional motion in which the skyrmion translates in one direction
under an oscillating applied electric field. 
The motion appears when there is a combination of two different
modulation frequencies
of the electric field which generate
rotation in and out of the plane of the director field 
on one side of the skyrmions \cite{Ackerman17,Duzgun18}.
Similar directed motion was also observed
in two-dimensional continuum
simulations \cite{Ackerman17}.
In theoretical studies
using coarse-grained models in which the skyrmions are represented
as solitons,
directed motion emerges
when the skyrmions move in a direction
perpendicular to the tilt of the 
background director,
but are unable to move as rapidly
in the opposite direction when the field is removed \cite{Long21}.

Skyrmions in chiral magnets
also undergo directed motion under an oscillating field,
and by superimposing
multiple driving fields, it is possible to achieve controlled
steering of the skyrmions
\cite{Moon16,Wang15,Yuan19,Chen19,Chen20}.
The directed motion in this case arises due to asymmetry in the oscillations
of the skyrmion shape,
and can be described in terms of a ratchet effect \cite{Reimann02}.
Ratchet effects also arise for the ac driving of skyrmions
coupled to an asymmetric substrate
\cite{Reichhardt15a,Ma17,Migita20,Gobel21,Souza21}.
For liquid crystal skyrmions, open questions include what ac driving protocols can
be used to produce directed motion,
such as whether a single ac driving frequency is sufficient to produce such
motion, and whether
reversals between forward and backward
motion or even multiple reversals
occur as the driving parameters are varied.
It is also interesting to explore whether ac drives that are not sinusoidal can
produce controlled directed motion.

In this work we examine the
ratchet-like
motion of a liquid crystal skyrmion under an oscillating ac field
with either a single driving frequency or multiple superimposed driving
frequencies.
We show that the skyrmion can translate in either the
forward or backward direction, and that multiple reversals of the motion
can occur as the
driving frequency  is varied.
We map out the different directed motion regimes for multiple frequency driving,
and show that it is possible to induce motion in each direction
using only
a single ac driving frequency.
The directed motion and reversals
also appear when we replace the sinusoidal driving by periodic square wave
pulses, which reveal more clearly that
the directed motion
results from different modes of skyrmion motion during
different portions of the driving cycle.
We discuss the relevance of our results
to other ratchet systems
in which directed motion
can occur in  the absence of an asymmetric substrate. 

\section{Numerical Methods}

\begin{figure}
  \begin{center}
    \includegraphics[width=0.8\columnwidth]{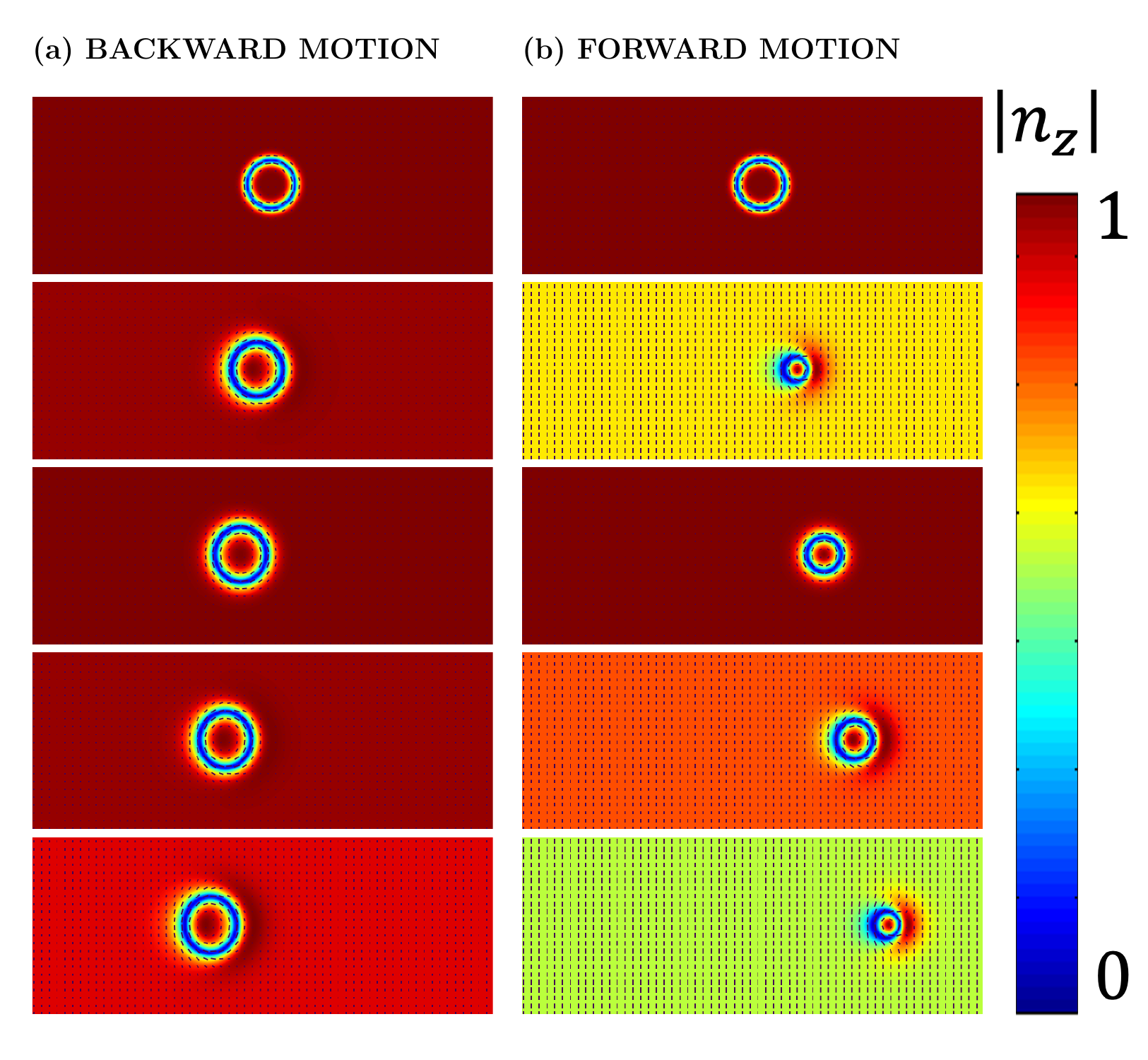}
    \end{center}
  \caption{
  The skyrmion positions and motion over time under ac driving.
  Color indicates the magnitude of the director field $|n_z|$ in the $z$
  direction, and time increases from top to bottom. (a)
	$A =0.1$, where the skyrmion translates in the negative $x$-direction
with $2\pi/\omega_1=1.2\times 10^{3}$ and $2\pi/\omega_2=1\times 10^4$. 
	(b) $A = 0.2$, where the skyrmion moves in the positive $x$-direction  
with $2\pi/\omega_1=1.2\times 10^{5}$ and $2\pi/\omega_2=1\times 10^{6}$. 	
	}
\label{fig:1}
\end{figure}

We consider a single liquid crystal skyrmion under two oscillating fields
using continuum based simulations of the type employed
previously
to model LC skyrmions \cite{Duzgun18,Duzgun20,Duzgun21}. 
In the continuum description, the traceless tensor $Q$ relates the
scalar order parameter $S$ to the orientational
order of a chiral nematic liquid crystal state,
which under proper constraints will support skyrmions
in systems confined between two substrates 
with normal surface anchoring.
The free energy density has the form
\begin{eqnarray}
  f &=& \frac{a}{2}\Tr(Q^2) + \frac{b}{3}\Tr(Q^3) + \frac{c}{4}[\Tr(Q^2)]^2\nonumber\\
  && + \frac{L}{2}(\partial_\gamma Q_{\alpha\beta})(\partial_\gamma Q_{\alpha\beta})
   - \frac{4\pi}{p}L\epsilon_{\alpha\beta\gamma}Q_{\alpha\rho}\partial_{\gamma}Q_{\beta\rho}\nonumber\\
  && - K[\delta(z) + \delta(z - N_z)]Q_{zz} -\Delta\epsilon  E^2\, \mathbf{\hat {n}}\cdot\mathbf{Q}\cdot\mathbf{\hat{n}}\ ,
\end{eqnarray}
where the nematic to isotropic transition is controlled by the terms 
$(a/2)\Tr(Q^2) + (b/3)\Tr(Q^3) + (c/4)[\Tr(Q^2)]^2$,
and the elastic energies with respect to a gradient in $Q$,
using the single elastic constant approximation,
are  $(L/2)(\delta_\gamma Q_{\alpha\beta})(\partial_\gamma Q_{\alpha\beta})
- (4\pi/p)L\epsilon_{\alpha\beta\gamma}Q_{\alpha\rho}\partial_\gamma Q_{\beta\rho}$,
which favor a twist with cholesteric pitch  $p$. 
The homeotropic surface anchoring
from the boundaries and the electric field
of magnitude $E$
along the unit vector ${\bf \hat n}$
arise in the last line with a coupling strength
$K$ and dielectric anisotropy $\Delta\epsilon$.
On the surfaces, the $Q$ tensor has uniaxial perfect ordering in the $z$ direction.
The electric field $E$ arises due to a potential difference across the slab. 
The skyrmion dynamics are obtained from the following overdamped equation:
$\partial_t Q({\bf r}, t) = -\Gamma\delta F/\delta Q({\bf r},t)$,
where $F=\int f({\bf r}) d^3 r$ and $\Gamma$ is the mobility constant.
As in previous work \cite{Duzgun18,Duzgun20,Duzgun21}
we employ $z$-invariant (2D) skyrmions \cite{Tai20}.
The out of plane rotation of skyrmions is generated by tilting the background
electric field.
This ac driving is produced by an $E$ field ${\bf E}=E [\sin(\theta){\bf \hat y} +\cos(\theta){\bf \hat z}]$ which
is periodically switched
between the positive $z$ direction and the positive
$y$ direction,
with polar angle $\theta = (\pi/6)[\cos(\omega_{1}t)\cos(\omega_{2}t)]^2$.

\section{Results}

In Fig.~\ref{fig:1} we show an image of the skyrmion under
ac driving, where the color code indicates the magnitude of the director
field $|n_z|$ in the $z$ direction.
In this case, when the electric field is tilted toward the positive $y$ axis, the
background director field is also tilted toward $y$ but the skyrmion shape is
deformed along the $x$ direction such that a crescent shape region of
vertical directors form on the positive $x$ side of the skyrmion. 
Here the electric field is oscillated between the $z$
direction and the positive $y$ direction. 
The skyrmion translates in the negative $x$ or backward
direction in Fig.~\ref{fig:1}(a),
where the parameters of the ac drive are
$2\pi/\omega_1=1.2\times 10^{3}$ and $2\pi/\omega_2=1\times 10^4$, 
while in Fig.~\ref{fig:1}(b), the same skyrmion
under ac driving with
$2\pi/\omega_1=1.2\times 10^{5}$ and $ 2\pi/\omega_2=1\times 10^{6}$
moves in the positive $x$ or forward direction.  
Crucially, in each case, an asymmetry in the skyrmion shape appears in the
orientation of the director field.

\begin{figure}
  \begin{center}
    \includegraphics[width=0.8\columnwidth]{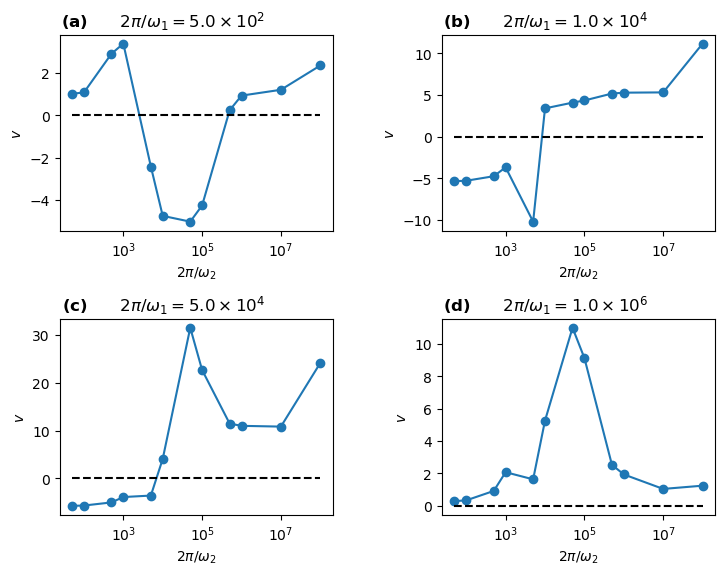}
    \end{center}
  \caption{
The skyrmion velocity $v$ versus $2\pi/\omega_{2}$  for the system 
in Fig.~\ref{fig:1} at $2\pi/\omega_{1} =$ (a) $5\times 10^2$,
(b) $1 \times 10^4$, (c)  $5\times10^4$,
and (d) $1 \times 10^6$.
There can be multiple reversals
in the direction of motion as a function of
frequency.
	}
\label{fig:2}
\end{figure}

We next hold $\omega_{1}$ fixed while varying
$\omega_2$, and measure the skyrmion velocity
over
a fixed number of ac drive cycles.
In Fig.~\ref{fig:2}(a) we plot the skyrmion velocity $v$ versus $2\pi/\omega_{2}$ for a
system with fixed $2\pi/\omega_{1} = 5\times 10^2$.
For
low frequencies,
the skyrmion translates in the positive $x$ direction,
but there a reversal to motion in
the negative $x$ direction for $1 \times 10^3 < 2\pi/\omega_{2} < 1 \times 10^5$,
followed by a second reversal to positive $x$ direction motion
for $2\pi/\omega_{2} > 1 \times 10^5$.
The magnitude of the
maximum velocity in the $-x$ direction is more than two times larger than the
magnitude of the maximum velocity in the $-y$ direction.

In Fig.~\ref{fig:2}(b) we plot the skyrmion velocity
for a sample with a much higher fixed frequency of
$2\pi/\omega_{1} = 1\times 10^4$.
For low values of $2\pi/\omega_2$,
the skyrmion moves
in the negative $x$ direction,
while a transition to motion in the positive $x$ direction
appears above
$2\pi/\omega_2=5\times 10^4$.
The overall magnitude of the motion
is much larger than that shown in
Fig.~\ref{fig:2}(a). 
For the sample with $2\pi/\omega_{1} = 5\times10^4$
in Fig.~\ref{fig:2}(c),
the velocity is weakly negative at low $2\pi/\omega_2$
and reverses to the positive $x$ direction for $2\pi/\omega_{2} > 1 \times 10^4$.
The magnitude of the maximum positive velocity is nearly six times larger than
the magnitude of the maximum negative velocity.
In Fig.~\ref{fig:2}(d) at $2\pi/\omega_{1} = 1 \times 10^6$,
the motion is only in the positive $x$ direction 
with a velocity peak near
$2\pi/\omega_{2}=1\times 10^{5}$.
The results in Fig.~\ref{fig:2} indicate that multiple reversals in the
direction of motion can occur as the frequency of the ac drive is varied.

\begin{figure}
  \begin{center}
    \includegraphics[width=0.8\columnwidth]{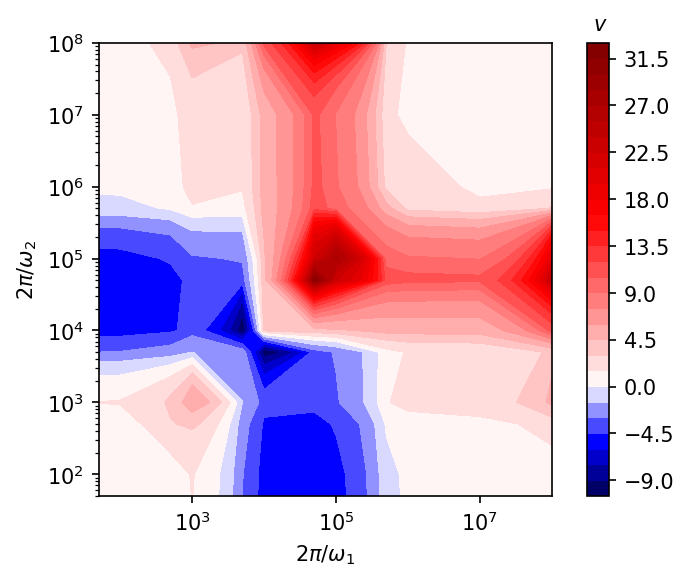}
    \end{center}
  \caption{
    Heat map of the skyrmion velocity $v$ in the negative $x$ (blue) and positive $x$
    (red) directions for the
    system in Fig.~\ref{fig:2} as a function of $2\pi/\omega_{2}$ versus $2\pi/\omega_{1}$.
    In some cases, multiple velocity reversals can occur as a function of changing
    frequency.
    Along the line $2\pi/\omega_{1} = 2\pi/\omega_{2}$, the system can be regarded as
    being driven by a single ac frequency, yet
    there are still ratchet reversals in the velocity
	response. 
	}
\label{fig:3}
\end{figure}

In Fig.~\ref{fig:3} we construct a skyrmion velocity map as a function of
$2\pi/\omega_{2}$ versus $2\pi/\omega_{1}$ for the system in
Figs.~\ref{fig:1} and \ref{fig:2}, where regions of positive and negative
direction motion appear along with regions in which no directed motion occurs.
Multiple velocity reversals can occur depending on the manner in which the
frequency is swept.
The greatest velocity magnitude occurs for positive $x$ direction motion
when $2\pi/\omega_{1} \approx 2\pi/\omega_{2} = 1 \times 10^5$.
When $2\pi/\omega_{2} > 1 \times 10^6$,
the motion is always in the positive $x$ direction.
Along the line $2\pi/\omega_{1} = 2\pi/\omega_{2}$, the 
system can be considered as being driven by a single frequency, and
even in this case, there are still
multiple reversals from positive to negative
velocity and back to positive velocity again.

\begin{figure}
  \begin{center}
    \includegraphics[width=0.6\columnwidth]{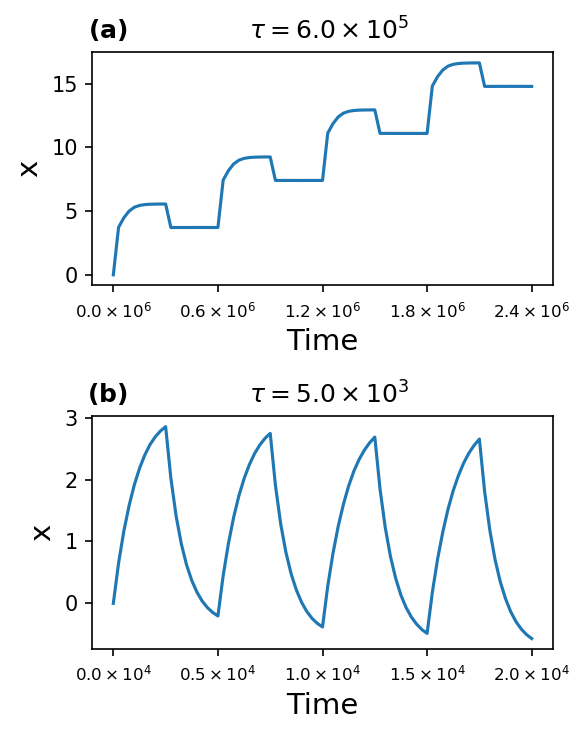}
    \end{center}
\caption{
The time series of the center of mass motion of the skyrmion under a 
square pulse drive with period
(a) $\tau=6\times 10^5$ , where the net motion is in the positive $x$ direction, and
(b) $\tau=5 \times 10^3$, where the net motion is in the negative $x$ direction.
The nature of the movement is different during the first portion of each
pulse compared to the second part of each pulse.
}
\label{fig:4}
\end{figure}

\begin{figure}
  \begin{center}
    \includegraphics[width=0.9\columnwidth]{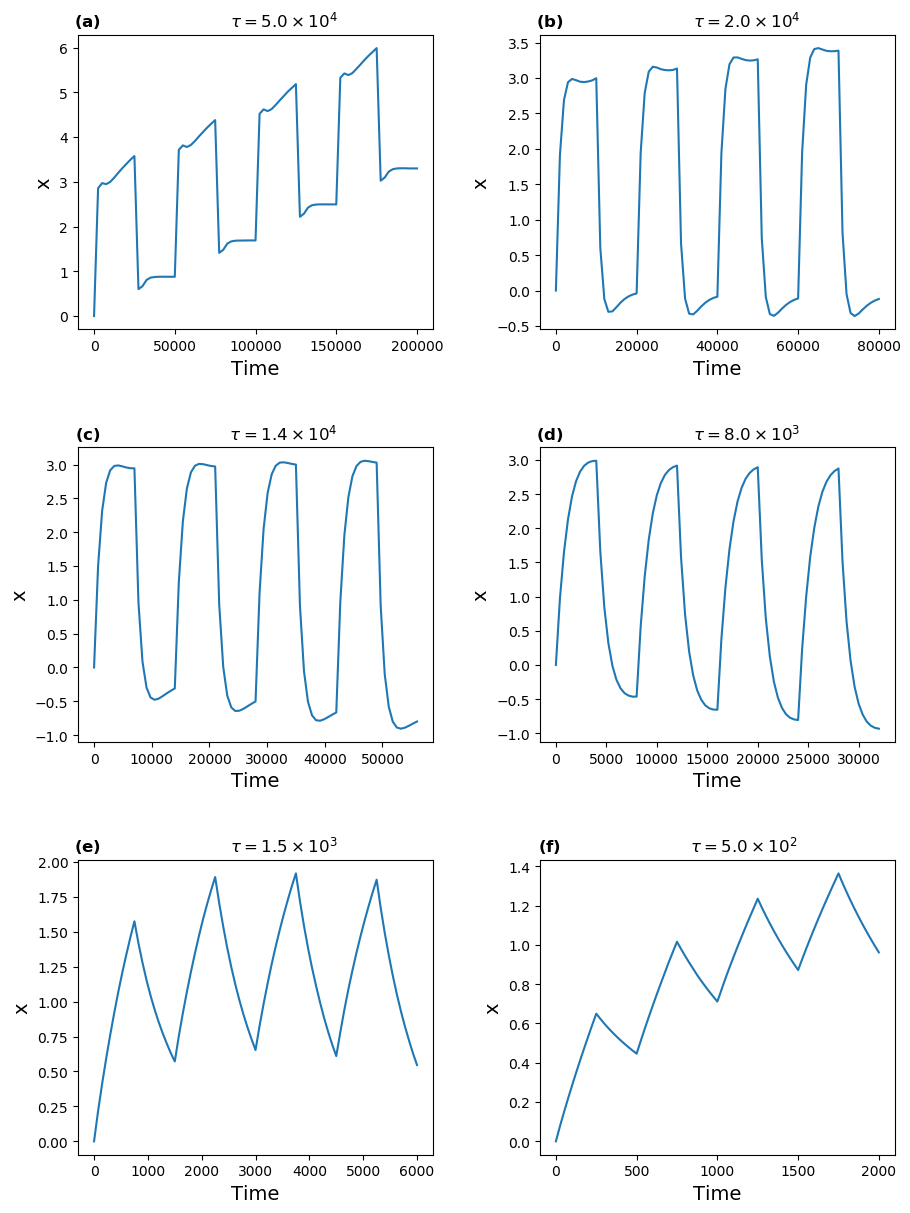}
    \end{center}
  \caption{
The time series of the skyrmion center of mass position $x$
for a system with a single periodic pulse drive of period $\tau$.
(a) At $\tau= 5\times10^4$ there is strong forward motion.
(b) At $\tau = 2\times10^4$, no directed motion occurs.
(c)  At $\tau=1.4\times10^4$, the motion is in the negative $x$ direction. 
(d)  Similar negative $x$ direction motion appears
for $\tau=8\times10^3$.
(e)  At $\tau=1.5\times 10^3$  there is no net motion.
(f) At $\tau=5\times 10^2$,
the motion is in the forward direction. 
	}
\label{fig:5}
\end{figure}

We next consider the effects of applying a periodic
pulse instead of a sinusoidal drive, which makes the transition
from positive to negative motion easier to distinguish.
To generate the periodic pulse, we set the $E$ field along
$\theta=30^\circ$ during the first half of each drive cycle, and then
set it to $\theta=0^\circ$ during the second half of each drive cycle.
In this case we consider only a single driving frequency.

In Fig.~\ref{fig:4} we plot the $x$ position of the skyrmion center of mass as
a function of time during four drive cycles in a sample where the
periodic square pulse drive
is applied with a period of $\tau=6\times 10^5$.
The skyrmion moves easily along the positive $x$ direction during the
$\theta = 30^\circ$ portion of the drive cycle, with the most rapid motion occurring
just after this field is first applied, while during the $\theta=0^\circ$
portion of the drive cycle, the skyrmion moves briefly backwards before stalling
and remaining stationary for the remainder of the drive cycle.
In Fig.~\ref{fig:4},
when the square pulse drive period is reduced to
$\tau=5 \times 10^3$,
the skyrmion has a slow net motion along the negative $x$ direction.
Here, there is a more rapid motion in the negative direction during the second
half of the driving cycle compared to the
more sluggish forward motion during the first half of
the driving cycle.

In Fig.~\ref{fig:5} we show a series of plots of the skyrmion
center of mass $x$ position versus time
for varied square pulse frequencies from high to low.
In Fig.~\ref{fig:5}(a), where the pulse period is $\tau= 5\times10^4$,
there is strong motion in the forward direction.
For $\tau=2\times10^4$
in Fig.~\ref{fig:5}(b),
no directed motion occurs.
In Fig.~\ref{fig:5}(c)
for $\tau=1.4\times10^4$,
there is a weak motion in the negative $x$ direction,
which becomes more prominent 
when $\tau=8\times10^3$ as shown in Fig.~\ref{fig:5}(d).
At $\tau = 1.5\times 10^3$  in Fig.~\ref{fig:5}(e) there is no net motion,
while for $\tau=5\times 10^2$, Fig.~\ref{fig:5}(f) indicates that
forward motion appears.
Here we find that
a series of reversals in the velocity
occur as a function of changing pulse frequency.

\begin{figure}
  \begin{center}
    \includegraphics[width=0.8\columnwidth]{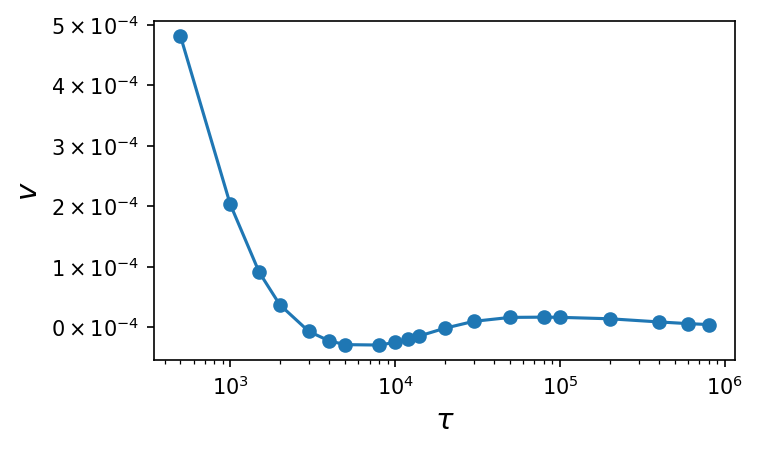}
    \end{center}
  \caption{
The skyrmion velocity $v$ versus period $\tau$ for the system
from Fig.~\ref{fig:5} with
a pulse drive
shows a series of reversals from positive
	to negative directed motion. 
	}
\label{fig:6}
\end{figure}

We plot the skyrmion velocity $v$ versus the pulse drive period $\tau$
in Fig.~\ref{fig:6},
where we find
a series of velocity reversals from positive motion for
periods
$\tau < 2\times 10^3$,
to negative motion, to a region of no  motion, and
finally to positive motion again.
For the highest pulse periods,
the skyrmion remains nearly static since the pulses become so rapid
that the director field is no longer able to respond to the ac driving.

\section{Discussion}
Our results indicate that multiple reversals in the directed motion
can occur for LC skyrmions under different ac drive conditions.
The behavior is similar
to the ratchet effect observed
in particle-like systems with an ac drive
when the particles are coupled to an asymmetric substrate
\cite{Reimann02}.
In the ratchet systems,
the particle moves along
the easy direction of the asymmetric substrate; however,
when multiple particles are present, collective interactions can induce
reversals or even multiple reversals of the direction of
ratchet motion
\cite{Reimann02,deSouzaSilva06a,Lu07,McDermott16}. 

It is also possible for ratchet effects to occur in the absence of an
asymmetric substrate
when some other form of asymmetry comes into play, such as
a nonlinear damping constant.
This can happen when particles are coupled to other particles, as
previously studied for superconducting vortices \cite{Cole06}
or assemblies of colloidal particles \cite{Reichhardt05a},
where the effective
damping of
each particle develops a velocity dependence
or a frequency dependence.
For example, if the ac drive is itself asymmetric, 
with a fast portion spanning a short time
$t_{1}$ with a large force $F_{1}$
and a slow portion spanning a longer time $t_{2}$ 
with a smaller force $F_{2}$,
such that $F_1t_1=-F_2t_2$, there is no net applied force during each driving
period of
$T = t_{1} + t_{2}$.   
A single particle with a fixed damping constant $\eta$
traverses a distance $d_{1} = (1/\eta) F_{1}t_{1}$ during the first portion of the
drive cycle and a distance
$d_{2} = (1/\eta )F_{2}t_{2}$
in the opposite direction during the second portion of the drive cycle.
If $\eta$ has no dependence on the drive,
then $d_1=d_2$ and there is no net ratcheting motion of the particle.
If, on the other hand, $\eta$ exhibits some 
nonlinear time dependence,
such that for high drive $F$ there is a shear thinning effect,
then
$d_{1}>d_2$
and the particle will move in the positive direction.
If instead there is shear
thickening, then
$d_{2}>d_1$ and net motion will occur in
the negative direction.
Effects of this type have been observed in a system
where each particle has a fixed damping coefficient
but the collective effects between particles in the surrounding medium
produce an effective nonlinear velocity 
dependence of the drag, leading to the emergence of a ratchet effect
\cite{Reichhardt05a}.

For the LC skyrmion system we consider here,
there is only a single skyrmion present,
but because the model is in 
the continuum limit,
collective modes can arise among the degrees of freedom.
Our results 
for the pulse drive suggest
that the net forward or backward motion originates
from a nonlinear drag effect
that is produced by asymmetric ac oscillations in the skyrmion shape.
Beyond LC skyrmions, the effects we observe could also be relevant for
skyrmions in magnetic systems
or for soft matter systems containing
bubble like shapes such as
vesicles undergoing some sort of asymmetric periodic
shape change or expansion \cite{Metselaar19}.
It would also be interesting to couple this motion to some kind of substrate 
in order to generate controlled directed motion. 

\section{Summary}
We have shown that liquid crystal skyrmions under an
ac
electric field
drive
biased along the positive $y$ direction
with either
multiple or single frequencies
can show directed motion in both the forward and backward directions
along the $x$ axis as
a function of frequency.
We map the dynamic regime diagram for this motion, showing 
that there are optimal frequencies for motion
and that multiple direction reversals can occur
even when only a single frequency is present.
We have also considered pulsed drives with a single frequency
and found that multiple reversal effects
can occur.
The ability to direct and precisely steer the motion of these deformable particles suggests that
this could be an interesting future route to the construction of 
soft robotic skyrmion systems.

\ack
This work was supported by the US Department of Energy through
the Los Alamos National Laboratory.  Los Alamos National Laboratory is
operated by Triad National Security, LLC, for the National Nuclear Security
Administration of the U. S. Department of Energy (Contract No. 892333218NCA000001).

\section*{References}
\bibliographystyle{iopart-num}
\bibliography{mybib}

\end{document}